\documentclass[10pt,conference]{IEEEtran}
\IEEEoverridecommandlockouts
\usepackage{bm}
\usepackage{cite}
\usepackage{amsmath,amssymb,amsfonts}
\usepackage{graphicx}
\usepackage{textcomp}
\usepackage{xcolor}
\usepackage{balance}
\usepackage{subfigure} 
\usepackage{amsmath}
\usepackage{subcaption}
\usepackage{algorithm}
\usepackage{caption}
\usepackage{geometry}
\usepackage{algorithm}
\usepackage{amsthm}

\newtheorem{proposition}{Proposition}

\usepackage{stfloats}
\newenvironment{proov} {{\it \noindent Proof. }} {\hfill $\blacksquare$\par}
\usepackage{algorithm}
\usepackage{algpseudocode}

\begin{document}
\captionsetup[figure]{labelformat={default},labelsep=period}
\title{Bayesian Beamforming for Integrated Sensing and Communication Systems}
\author{Zongyao Zhao$^{1,2}$, Zhenyu Liu$^{1}$, Wei Dai$^{3}$, Xinke Tang$^{2}$, Xiao-Ping Zhang$^{1}$, Yuhan Dong$^{1,2,*}$\\
	$^1$Shenzhen International Graduate School, Tsinghua University, Shenzhen, P. R. China\\
    $^2$Pengcheng Laboratory, Shenzhen, P. R. China\\
    $^3$Department of Electrical and Electronic Engineering, Imperial College London, UK\\
	Email: zhaozong21@mails.tsinghua.edu.cn, zhenyuliu@sz.tsinghua.edu.cn, wei.dai1@imperial.ac.uk\\
              tangxk@pcl.ac.cn, 
              xiaoping.zhang@sz.tsinghua.edu.cn,
                dongyuhan@sz.tsinghua.edu.cn}

 \maketitle

\begin{abstract}
The uncertainty of the sensing target brings great challenge to the beamforming design of the integrated sensing and communication (ISAC) system. To address this issue, we model the scattering coefficient and azimuth angle of the target as random variables and introduce a novel metric, expected detection probability  (${EP_d}$), to quantify the average detection performance from a Bayesian perspective. Furthermore, we design a Bayesian beamforming scheme to optimize the expected detection probability under the limited power budget and communication performance constraints. A successive convex approximation and semidefinite relaxation based (SCA-SDR) algorithm is developed for the complicated non-convex optimization problem corresponding to the beamforming scheme. Simulation results show that the proposed scheme outperforms other benchmarks and exhibits robust detection performance when parameters of the target are unknown and random.
\end{abstract}


\section{Introduction}
Integrated sensing and communication (ISAC) is one of the pivotal technologies for the next generation of mobile communications\cite{Cui2021,Hassan2016,Liu2022}. By sharing wireless spectrum resources, ISAC enables traditional network infrastructure to acquire sensing capabilities in a low-cost manner. This integration of sensing and communication capabilities is expected to enable many promising emerging applications \cite{Zhao2022,LiuF2024,Kwon2023}, such as smart cities, autonomous driving, etc.

Beamforming technology has always been an important research topic in the ISAC area\cite{ZhaoZ2024,LiuX2020,LiuF2022,Zhao2024,Zhao2025}. By carefully designing the beamformer, the communication and sensing performance of the ISAC system can be effectively enhanced. Extensive research on beamforming design has been conducted in both academia and industry. For instance, the author in \cite{LiuX2020} proposed a joint beamforming scheme to optimize the beampattern under the constraint of communication signal-to-interference-plus-noise ratio (SINR). In \cite{LiuF2022} the authors developed a scheme for Cram\'er-Rao bound (CRB) optimization under the constraint of communication SINR. A joint beamforming design is proposed in \cite{Zhao2024} for robust CRB optimization in multi-target scenarios. The authors in \cite{Zhao2025} proposed a joint beamforming scheme to optimizie multi-target detection performance in multi-target scenarios.

Most of existing works assume that the target parameter is determined and known in advance, and design the beamformer to optimize the performance of communication and sensing. However, in practice, targets are always unknown and random. Although we can't know the true value of the parameter, we may be able to obtain the prior distribution of the parameter from historical data or motion models. Several works have been carried out to explore this case. The authors in \cite{AttiahX2023} proposed an active beamforming scheme, using the posterior distribution of parameters for beamforming design. In \cite{Xu2024}, the authors modeled the target parameter as a random variable, derived the upper bound of the posterior CRB (PCRB), and proposed a transmit covariance matrix optimization scheme to minimize the sensing PCRB. In \cite{Hou2024}, the authors propose a secure ISAC beamforming scheme aimed at maximizing the worst-case secrecy rate across all possible target locations, subject to a sensing PCRB threshold. However, in these works, performance indicators such as expected SINR and PCRB are considered. Research on the performance of target detection from the Bayesian expectation perspective remains scarce.

In this paper, we address the challenge of target uncertainty and focus on the scenario where the parameters are unknown and random case. We model the scattering coefficient and azimuth angle of the target as random variables with known distributions. Moreover, we define a new sensing metric called expected detection probability  ${EP_d}$ to measure the average detection performance from a Bayesian perspective. Then, we formulate the Bayesian beamforming scheme as a ${EP_d}$ optimization problem under the communication performance and power budget constraints, and develop a successive convex approximation and semidefinite relaxation based (SCA-SDR) algorithm to solve the non-convex problem. The simulation results demonstrate the effectiveness of the proposed scheme.

\emph{Notation}: In this paper, Bold italic lower-case and upper-case letters denote vectors and matrices respectively. $\mathbb{R}$ and $\mathbb{C}$ represent the real and complex sets respectively. $|\cdot|$, $||\cdot||$, and $||\cdot||_{F}$ are  absolute value, Euclidean norm, and Frobenius norm, respectively. $\left( \cdot \right)^{-1}$ and $\left( \cdot \right)^{\dagger}$ denote the inverse and pseudo inverse, respectively. $\left( \cdot \right)^T$, $\left( \cdot\right)^*$, and $\left( \cdot \right)^H$ represent transpose, complex conjugate, and Hermitian transpose, respectively. $\mathbb{E}\left( \cdot \right)$ represents statistical expectation. $\mathrm{Re}\left\{ \cdot\right\}$ returns the real part of a complex number. $j$ is the imaginary unit, which means $j^2=-1$. $\mathbf{I}_N$ is the $N\times N$ identity matrix. $\mathbf{1}=\left[1,1,\ldots,1\right]^T\in\mathbb{R}^N$. $\bm{A}\succeq0$ means that $\bm{A}$ is a positive semidefinite matrix. $\odot$ represents the Hadamard product. $\mathrm{diag}\left(\bm{a} \right)$ returns a diagonal matrix, the vector composed of its diagonal elements is $\bm{a}$. $\mathrm{Tr}\left(\bm{A} \right)$ and $\mathrm{rank}\left(\bm{A}\right)$ compute the trace and rank of matrix $\bm{A}$ respectively. $\mathrm{chol}\left(\bm{A}\right)$ returns the Cholesky decomposition of matrix $\bm{A}$. $\mathrm{vec}\left(\bm{A}\right)$ vectorize matrix $\bm{A}$ by column-stacking.

The remainder of this paper is organized as follows. Sect.~\ref{sec2} introduces the signal and system model. We propose the Bayesian beamforming scheme in Sect.~\ref{sec3} and present numerical results in Sect.~\ref{sec4}. Finally, the conclusions are drawn in Sect.~\ref{sec5}.

\section{System and Signal Model} \label{sec2}
\subsection{System Model}
\begin{figure}[t]
	\centering
\includegraphics[width=0.48\textwidth]{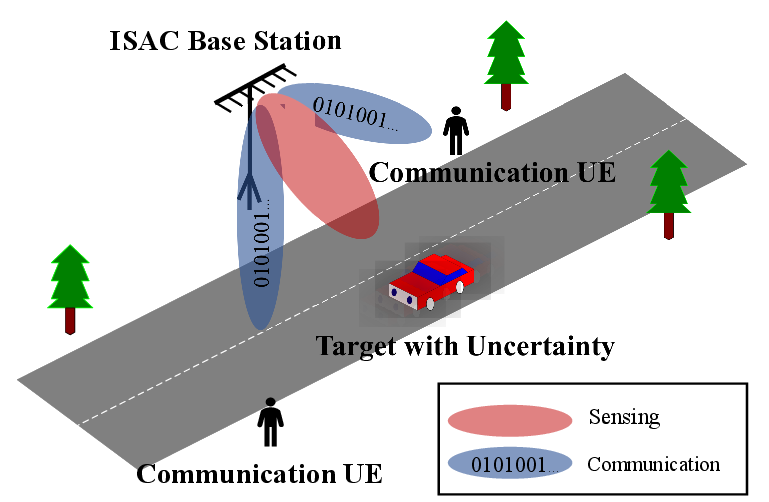}\\	\caption{Illustration of the ISAC system with an uncertain target.}
        \label{Fig1}
\end{figure}
We consider an ISAC base station with $N_t$ transmit antennas and $N_r$ receive antennas intending to provide downlink communication services for $K$ single antenna user equipment (UEs) and sensing a target with uncertainty simultaneously, as shown in Fig~\ref{Fig1}. The transmit signal $\bm{X}$  can be expressed as
\begin{gather}
\label{eq1}
\begin{aligned}
\bm{X}=\bm{W}_c\bm{S}_c+\bm{W}_s\bm{S}_s=\bm{WS},
\end{aligned}
\end{gather}
where $\bm{W}_c=[\bm{w}_1, \bm{w}_2,\ldots, \bm{w}_K]\in \mathbb{C} ^{N_t\times K}$ and $\bm{W}_s\in\mathbb{C} ^{N_t\times N_t}$ are communication and sensing beamformer, respectively. $\bm{S}_c\in \mathbb{C} ^{K\times L}$ and $\bm{S}_s\in \mathbb{C} ^{N_t\times L}$ are communication data and a dedicated sensing waveform of length $L$. So, $\bm{W}=\left[ \bm{W}_c,\bm{W}_s \right] \in \mathbb{C} ^{N_t\times \left( N_t+K \right)}$ and $\bm{S}=\left[ \bm{S}_c,\bm{S}_s \right]^{T} \in \mathbb{C} ^{(N_t+K)\times L}$ are joint beamformer and joint
data augmentation matrix. When signal length is sufficiently large, it is assumed that there is no correlation between dedicated sensing waveforms and random communication data. The sample covariance matrix of $\bm{X}$ is given by \cite{LiuF2022}
\begin{gather}
\label{eq2}
\begin{aligned}
\bm{R}_{\bm{X}}=\frac{1}{L}\bm{XX}^H\approx\bm{WW}^H\in \mathbb{C} ^{N_t \times N_t}.
\end{aligned}
\end{gather}
\subsection{Sensing Model}
The echo signals at the ISAC base station are denoted by 
\begin{align}\label{sensing_signals}
    \bm{Y}_{s}& = \bm{H}_s \bm{X} + \bm{Z}_s,
\end{align}
where $\bm{H}_s={\alpha}
\bm{b}\left( \mathrm{\theta} \right) \bm{a}^H\left( \mathrm{\theta} \right)=\alpha\bm{A}\left( \mathrm{\theta} \right) \in \mathbb{C}^{N_r \times N_t}$ is the target response matrix. $\alpha$ and $\theta$ are the complex scattering coefficient and azimuth angle of the target. $\bm{a}\left( \mathrm{\theta} \right)$ and $\bm{b}\left( \mathrm{\theta} \right)$ are steering vectors of the transmit and receive array. $\bm{Z}_s \in \mathbb{C}^{N_r \times L}$ denotes an additive noise matrix with each entry following complex Gaussian distribution $\mathcal{CN}(0,\sigma_s^2)$. Since the position of the target is uncertain, we model $\alpha$ and $\theta$ as independent random variables.

By defining $\bm{V}\left( \theta \right) = \bm{I}_L\otimes\left(\bm{b}\left( \mathrm{\theta} \right) \bm{a}^H\left( \mathrm{\theta} \right) \right)$, and $\bm{x}=\mathrm{vec}\left(\bm{X}\right)$, $\bm{z}_s=\mathrm{vec}\left(\bm{Z}_{s}\right)$, the $\bm{y}_s$ can be rewritten as
\begin{gather}
\begin{aligned}
\bm{y}_s \triangleq  \mathrm{vec}\left( \bm{Y}_s \right)= {\alpha}\bm{V}\left( \mathrm{\theta}\right)\bm{x}+\bm{z}_s.
\label{eq4}
\end{aligned}
\end{gather}
We consider the detection task of the target, which can be easily formulated as a hypothesis testing problem as follows
\begin{align}
\begin{cases}
	\mathcal{H} _0:\bm{y}_{s} =  \bm{z}_s\\
	\mathcal{H} _1:\bm{y}_{s} = {\alpha}\bm{V}\left( \mathrm{\theta}\right)\bm{x}+\bm{z}_s,
\end{cases}
\label{eq5}
\end{align}
\normalsize
where $\mathcal{H} _0$ means there is no target echo signal, $\mathcal{H} _1$ means there is target echo. Since both $\alpha$ and $\theta$ are unknown parameters, the generalized likelihood ratio test (GLRT) can be used for target detection.
\begin{proposition}
In this case, the generalized likelihood ratio test (GLRT) detector can be expressed as, 
\begin{align}
\max_{\theta} \frac{\left|\bm{y}_s^H \bm{V}\left( \theta\right)\bm{x}\right|^2}{\|\bm{V}\left( \theta\right)\bm{x}\|^2} \begin{array}{c}
	\overset{\mathcal{H}_1}{\geqslant}\\
	\underset{\mathcal{H}_0}{<}\\
\end{array}\,\, \eta,
\label{eq6}
\end{align}
\label{lm0}
where $\bm{y}_s$ is the received signal, $\bm{x}$ is the transmitted signal, and $\eta$ is the detection threshold. The detector achieves constant false alarm rate (CFAR) performance by controlling the threshold $\varGamma$.
\end{proposition}
\vspace{-0.5em}
\begin{proov}
See Appendix~\ref{AP1}.
\end{proov}

In an actual system, it is usually impossible to know at what angle the detection statistic obtains the maximum value. More practical, the scheme usually adopted is to perform separate detections at multiple angle grid points. For the angle $\theta_j$ , we can get the following detector \cite{Tang2022,Neyman1992,Kay1998,Fortu2020}
\begin{gather}
\begin{aligned}
\label{eq7}
\mathrm{Re}\left
\{\bm{y}_s^H \bm{V}\left( \theta_j\right)\bm{x}\right\} \begin{array}{c}
	\overset{\mathcal{H}_1}{\geqslant}\\
	\underset{\mathcal{H}_0}{<}\\
\end{array}\,\, \varGamma,
\end{aligned}
\end{gather}
and $\varGamma$ is the detection threshold. Since the angle to be detected is completely known at this time, the denominator of the GLRT statistic becomes a constant, and the GLRT detection is completely degenerated into the above matched filter detector. The detection probability of the target at angle $\theta_j$ is denoted as
\begin{flalign}\label{eq8}
\noindent P_d(|\alpha|,\theta_j)=\frac{1}{2}\mathrm{erfc}\left( \mathrm{erfc}^{-1}\left(2P_f\right) -\sqrt{\frac{|\alpha|^2\mathrm{tr}\left( \boldsymbol{F}\left( \theta_j \right) \boldsymbol{R}_{\boldsymbol{X}} \right)}{\sigma _{s}^{2}}} \right)
\end{flalign}
where $\mathrm{erfc}\left( x \right) =\frac{2}{\sqrt{\pi}}\int_x^{\infty}{e^{-t^2}}dt$ is  the complementary error function, the constant $P_{f}$ is the probability of false alarm, and $\boldsymbol{F}\left( \theta \right)=\bm{A}\left( \theta \right)\bm{A}^{H}\left( \theta \right)$. 

In this work, we consider that $\alpha$ and $\theta$ are random variables, we define a new expected detection probability  ($EP_d$) metric, measuring the average detection performance, which is given by
\begin{align}\label{eq9}
EP_d&=\mathbb{E} _{|\alpha| ,\theta}\left( P_d(|\alpha|,\theta) \right) =\iint{f\left( \theta \right) f\left( |\alpha| \right) P_d(\alpha,\theta)d|\alpha|d\theta}  
\end{align}
where $f\left( |\alpha| \right)$ and $f\left( \theta \right)$ are the probability density function of the random variable $|\alpha|$ and $\theta$. 

\subsection{Communication Model}
The received signal by the $k$-th  UE can be expressed as
\begin{align}\label{eq10}
    \bm{y}_{c}^{k}& = \bm{h}_k^{H} \bm{X}  + \bm{z}_c,
\end{align}
where $\bm{h}_k^{H}\in \mathbb{C} ^{1\times N_t}$ is the channel from the transmit antenna array to the $k$-th UE for $k=1,2,\ldots, K$. $\bm{z}_{c}\sim \mathcal C\mathcal N \left( 0,\sigma_{c}^{2}\bm{I}_L \right)$ is the communication noise. We assume that the communication channel experience both small-scale and large-scale fading. The channel $\bm{h}_k^{H}$ can be expressed as \cite{Tse2005}
\begin{gather}
\label{eq11}
\begin{aligned}\boldsymbol{h}_{k}^{H}=\sqrt{\eta _k}\left( \sqrt{\frac{\kappa}{\kappa +1}}\bar{\boldsymbol{g}}_k+\sqrt{\frac{\kappa}{\kappa +1}}\tilde{\boldsymbol{g}}_k \right),\end{aligned}
\end{gather}
where $\eta _k$ is the large-scale fading coefficient and $\sqrt{\frac{\kappa}{\kappa+1}}\bar{\boldsymbol{g}}_k+\sqrt{\frac{\kappa}{\kappa +1}}\tilde{\boldsymbol{g}}_k$ is the small-scale Rician fading model. $\eta _k$ is modeled as $-20\log _{10}\left( \lambda /\left( 4\pi d_0 \right) \right) +10n\log _{10}\left(d_k/d_0 \right)$ (in dB), where $d_k$ is the distance between transmit array and $k$-th UE, $d_0=1$ m is reference distance, $n=2.2$ is the path loss exponent, and $\lambda$ is the wavelength of the carrier frequency. in small-scale fading part, $\kappa$  is the energy ratio between the line-of-sight (LoS) path and none-LoS (NLoS) paths, $\bar{\boldsymbol{g}}_k=\bm{a(\theta)}$ is the LoS channel, and $\tilde{\boldsymbol{g}}_k\sim \mathcal C\mathcal N \left( 0,\bm{I}_{N_t} \right)$ is the NLoS channel. 
 
The communication SINR of the $k$-th UE is given by
\begin{gather}
\begin{aligned}
\label{eq12}
\gamma _k=\frac{|\bm{h}_{k}^{H}\bm{w}_k|^2}{\,\, \sum_{i=1,i\ne k}^{i=K}{|\bm{h}_{k}^{H}\bm{w}_i|^2}+||\bm{h}_{k}^{H}{\bm{W}_s}||^2+\sigma _{c}^{2}},
\end{aligned}
\end{gather}
where $\sum_{i=1,i\ne k}^{i=K}{|\bm{h}_{k}^{H}\bm{w}_i|^2}$ is the multi-user interference. $||\bm{h}_{k}^{H}\bm{W}_s||^2$ is the interference caused by the dedicated sensing signal.

\section{Bayesian Beamforming Scheme} \label{sec3}
In order to improve the expected detection probability, we design a novel beamforming scheme in this section. We formulate the beamforming design scheme as the following optimization problem,
\begin{subequations}\label{eq13}
\begin{align}
\left( \mathcal{P} _1 \right)~~  & \mathop{\mathrm{maximize}} \limits_{\,\,\bm{W}}   &&EP_d 
\\
&~\mathrm{subject~to}   &&\gamma_{k}\geqslant \gamma _{{\mathrm{th}}},~\forall k \label{eq13b}\\
& &&\mathrm{tr}\left( \bm{WW}^H \right) \leqslant P_T, \label{eq13c}
\end{align}
\end{subequations}
where beamformer $\bm{W}$ is the optimization variable, \eqref{eq13b} is to set a threshold for the communication SINR of each UE, and \eqref{eq13c} is a power budget. Introducing auxiliary variables $\bm{R}_{\bm{X}}=\bm{WW}^{{H}}$,  $\bm{W}_k=\bm{w}_k\bm{w}_{k}^{H}$, and $\bm{Q}_k=\bm{h}_{k}^{*}\bm{h}_{k}^{T}$, then \eqref{eq11} can be reformulated as 
\begin{gather}
\begin{aligned}
\label{eq14}
\gamma _k=\frac{\mathrm{tr}\left( \bm{Q}_k\bm{W}_k \right)}{\mathrm{tr}\left( \bm{Q}_k\left( \bm{R}_\mathbf{X}-\bm{W}_k \right) \right) +\sigma _{c}^{2}}.
\end{aligned}
\end{gather}
\setcounter{equation}{16}
\begin{figure*}[t]
\begin{align}\label{eq17}
&EP_d \approx \mathcal{F} \left( \boldsymbol{R}_{\boldsymbol{X}}^{\left( t \right)} \right) =\sum_{m=1}^M\sum_{n=1}^N{w_{\theta}}\left( \theta _m \right) {w_{\alpha}}\left( \alpha _n \right) \left[ P_d\left( \alpha _n,\theta _m,\boldsymbol{R}_{\boldsymbol{X}}^{\left( t \right)} \right)
+\left< \nabla P_d\left( \alpha _n,\theta _m,\boldsymbol{R}_{\boldsymbol{X}}^{\left( t \right)} \right) ,\left( \boldsymbol{R}_{\boldsymbol{X}}-\boldsymbol{R}_{\boldsymbol{X}}^{\left( t \right)} \right) \right> \right] \\
&\nabla P_d\left( \alpha _n,\theta _m,\bm{R}_{\bm{X}}^{\left( t \right)} \right) =\frac{2|\alpha _n|^2\boldsymbol{F}^T\left( \theta _m \right)}{\sqrt{\pi}\sigma _{s}^{2}\sqrt{\frac{|\alpha|^2\mathrm{tr}\left( \boldsymbol{F}\left( \theta_m \right) \boldsymbol{R}_{\boldsymbol{X}}^{\left( t \right)} \right)}{\sigma _{s}^{2}}}}\exp \left( -\left( erfc^{-1}\left( 2P_f \right) -\sqrt{\frac{|\alpha_n|^2\mathrm{tr}\left( \boldsymbol{F}\left( \theta_m \right) \boldsymbol{R}_{\boldsymbol{X}}^{\left( t \right)} \right)}{\sigma _{s}^{2}}} \right) ^2 \right) \label{eq18}
\end{align}\normalsize 
\hrulefill
\end{figure*}
\setcounter{equation}{14}

Next, substituting auxiliary variables \eqref{eq14} into \eqref{eq13}, the Problem $\left( \mathcal{P} _1 \right)$ can be transformed as
\begin{subequations}\label{eq15}
\begin{align}
\left( \mathcal{P} _{1.1} \right)~~  & \mathop{\mathrm{maximize}} \limits_{\,\,\bm{R}_{\bm{X}},\bm{W}_{\mathrm{1}},...,\bm{W}_{K}}   &&EP_d
\\
&~~~\mathrm{subject~to}   &&\mathrm{tr}\left( \bm{Q}_k\bm{W}_k \right)-\gamma_\mathrm{th}\mathrm{tr}\left( \bm{Q}_k\left( \bm{R}_{\bm{X}}-\bm{W}_k \right) \right) \label{eq15b}\nonumber\\
& &&  \geqslant \gamma_\mathrm{th} \sigma _{c}^{2}, ~\forall k  
\\
& &&\mathrm{tr}\left( \bm{R}_{\bm{X}} \right) \leqslant P_T
\\
& && \bm{R}_{\bm{X}}-\sum_{k=1}^{K}\bm{W}_{k}\succeq 0
\\
& &&\mathrm{rank}\left( \bm{W}_k \right) =1, \forall k. \label{eq15e}
\end{align}
\end{subequations}
According to \eqref{eq8} and \eqref{eq9}, the objective function is a non-convex function that contains nonlinear function $\mathrm{erfc}\left( x \right)$ and integrals. The integration is a tricky operation for optimization problems. Here, we discretize the distribution and then use the summation operation instead of the complicated integration. As a result, \eqref{eq8} is approximated as
\begin{align}
\label{eq16}
EP_d \approx \sum_{m=1}^M\sum_{n=1}^N{w_{\theta}}\left( \theta _m \right) {w_{\alpha}}\left( \alpha _n \right)P_d(\alpha_n,\theta_m), 
\end{align}
where ${w_{\theta}}\left( \theta _m \right)$ and ${w_{\theta}}\left( \alpha_n \right)$ are probability values of the corresponding values of $\theta_m$  and $\alpha_n$ in the discretized distribution, and $\sum_{m=1}^M{w_{\theta}}\left( \theta _i \right) =\sum_{n=1}^N{w_{\alpha}}\left( \alpha _n \right) =1$.
\begin{algorithm}[t]
    \caption{SCA-SDR Algorithm for Solving $\left( \mathcal{P} _{1} \right)$.}
    \label{Alg1}
    \begin{algorithmic}[1]
    \Require
    $P_T$, $\sigma_{c}^2$, $\sigma_{s}^2$, $\bm{h}_1^{H}$, $\bm{h}_{2}^{H}$,..., $\bm{h}_{K}^{H}$, $N_t$, $N_r$, $\gamma_{k}$, $\mu_{\theta}$, $\sigma_{\theta}^{2}$, $\mu_{\alpha}$, $\sigma_{\alpha}^{2}$, $M$, $N$.
    \Ensure
    $\bm{W}^{*}$.
    \State Initialize: $\bm{W}^{(t)}, t = 1$, $\varepsilon_\mathrm{th}$, $\varepsilon=+\infty$, ${T}_{\text{max}}$.
    \While{$t \leqslant {T}_{\text{max}}$ and  $\varepsilon \geqslant \varepsilon_\mathrm{th}$ }
    \State Calculate  $\bm{R}_{\bm{X}}^{(t)}=\bm{W}^{(t)}{\bm{W}^{H(t)}}$ 
    \State Calculate $\mathcal{F} \left( \boldsymbol{R}_{\boldsymbol{X}}^{\left( t \right)} \right)$ via \eqref{eq17} and \eqref{eq18}
    \State Obtain $\overline{\bm{R}}_{\bm{X}},\overline{\bm{W}}_{1},\ldots,\overline{\bm{W}}_{K}$ by solving \eqref{eq19} without rank-1 constraints via CVX toolbox
    \State Calculate  $\widetilde{\bm{R}}_{\bm{X}},\widetilde{\bm{W}}_{1},\ldots,\widetilde{\bm{W}}_{K}$ via \eqref{eq20}\eqref{eq21}
    \State Compute $\bm{W}^{\dagger}$ via \eqref{eq22}\eqref{eq23}\eqref{eq24}
    \State Update $\bm{W}^{(t+1)}=\bm{W}^{(t)}+\delta^{(t)}(\bm{W}^{\dagger}-\bm{W}^{(t)})$\normalsize, where $\delta^{(t)}$ is the stepsize calculated using the Armijo rule
    \State Update $\varepsilon=\delta^{(t)}\left\| \bm{W}^{\dagger}-\bm{W}^{(t)} \right\|_F$
    \State Update $t=t+1$
    \EndWhile \\
    \Return $\bm{W}^{*}=\bm{W}^{(t)}$.
    \end{algorithmic}
\end{algorithm}

Note that the objective function \eqref{eq15} is still nonlinear due to the presence of the $\mathrm{erfc}\left( x \right)$ function. We propose to tackle this by the successive convex approximation (SCA) algorithm \cite{{Scutari2014}}. Given a feasible point $\boldsymbol{W}^{\left( t \right)}$, we have $\boldsymbol{R}_{\boldsymbol{X}}^{\left( t \right)}=\bm{W}^{(t)}{\bm{W}^{H(t)}}$, the first
order Taylor convex approximation at $\boldsymbol{R}_{\boldsymbol{X}}^{\left( t \right)}$ is given by \eqref{eq16}, where the $\left< \bm{A},\bm{B}\right>$ is the inner product of $\bm{A}$ and $\bm{B}$, and $\nabla P_d\left( \alpha _n,\theta _m,\boldsymbol{R}_{\boldsymbol{X}}^{\left( t \right)} \right)$ is the gradient of $EP_d$ calculated by \eqref{eq17}. As a result, at the $t$-iteration of the SCA algorithm, we solve the following optimization problem
\setcounter{equation}{18}
\begin{align}\label{eq19}
\left( \mathcal{P} _{1.2} \right)~~  & \mathop{\mathrm{maximize}} \limits_{\,\,\bm{R}_{\bm{X}},\bm{W}_{\mathrm{1}}, ...,\bm{W}_{k}}   &&\mathcal{F} \left( \boldsymbol{R}_{\boldsymbol{X}}^{\left( t \right)} \right)
\\
&~~~\mathrm{subject~to}   &&\eqref{eq15b}-\eqref{eq15e}. \nonumber
\end{align}
The above problem is a typical semidefinite relaxation (SDR) problem \cite{Luo2010}. Ignoring the rank-1 constraint \eqref{eq15e}, $\left( \mathcal{P} _{1.2} \right)$ is a convex problem, which can be solved using the CVX tool \cite{boyed2004} to obtain the optimal solution $\overline{\bm{R}}_{\bm{X}},\overline{\bm{W}}_{1},\ldots,\overline{\bm{W}}_{k}$. Further using the following proposition, the optimal rank-1 solution can be obtained directly \cite{LiuX2020}.
\begin{proposition}
Given an optimal solution $\overline{\bm{R}}_{\bm{X}},\overline{\bm{W}}_{1},\ldots,\overline{\bm{W}}_{k}$ of $(\mathcal{P} _{1.2})$ without rank-1 constrains, the following $\widetilde{\bm{R}}_{\bm{X}},\widetilde{\bm{W}}_{1},\ldots$\\ $,\widetilde{\bm{W}}_{k}$ is the rank-1 optimal solution of $(\mathcal{P} _{1.2})$
\begin{align}
&\widetilde{\bm{R}}_{\bm{X}}=\overline{\bm{R}}_{\bm{X}},\label{eq20}\\ &\widetilde{\bm{W}}_{k}=\frac{\overline{\bm{W}}_{k}\bm{Q}_{k}\overline{\bm{W}}_{k}^{H}}{\mathrm{tr}\left( \bm{Q}_{k}\overline{\bm{W}}_{k} \right)}, \forall k.\label{eq21}
\end{align}
\label{lm1}
\end{proposition}
\vspace{-0.5em}
\begin{proov}
See \cite[Theorem 1]{LiuX2020}.
\end{proov}
Next, The optimal beamforming matrix solution $\bm{W}^{\dagger}$ at $t$-iteration step of SCA can be obtained by
\begin{align}
&\widetilde{\bm{w}}_{k}=\left( \bm{h}_{k}\widetilde{\bm{W}}_{k} \bm{h}_{k}^{H} \right) ^{-1/2}\widetilde{\bm{W}}_{k}\bm{h}_{k}^{H}, \forall k, \label{eq22}
\\
&\widetilde{\bm{W}}_{s}=\mathrm{chol}\left(\widetilde{\bm{R}}_{\bm{X}}-\sum_{k=1}^{K}\widetilde{\bm{W}}_{k}\right),\label{eq23}
\\
&\bm{W}^{\dagger}=\left[\widetilde{\bm{w}}_{1},\ldots \widetilde{\bm{w}}_{K}, \widetilde{\bm{W}}_{s} \right],\label{eq24}
\end{align} 
where $\mathrm{chol}$ is the Cholesky decomposition. Next, we update the $\bm{W}^{(t+1)}=\bm{W}^{(t)}+\delta^{(t)}(\bm{W}^{\dagger}-\bm{W}^{(t)})$, where $\delta^{(t)}$ is the stepsize obtained from Armijo rule to ensure the convergence. we iteratively update the optimization variable until convergence. The proposed algorithm for solving $\left( \mathcal{P} _{1} \right)$  is summarized in Algorithm \ref{Alg1}. The total complexity of Algorithm 1 is $\mathcal{O} \left( I_{iter}K^{4.5}N_{t}^{4.5}\log \left( 1/\varepsilon \right) \right)$, where  $I_{iter}$ is the iterations required for the SCA algorithm to converge, $\varepsilon$ is the accuracy of interior-point method.

\section{Numerical Results} \label{sec4}
We present numerical results of the proposed Bayesian beamforming scheme in this section. We set the transmit and receive array with the same number of antennas, i.e., $N_t =N_r=16$, the transmit power as $P_T=20$dBm, and the noise power of communication and sensing as $\sigma _{c}^{2}=\sigma _{s}^{2}=-94$dBm. The carrier frequency of the system is set to $f_c=2.4$ GHz. We consider $K=2$  communication UEs at $(-45^{\mathrm{o}}, 200\mathrm{m})$ and $(45^{\mathrm{o}}, 200\mathrm{m})$. With $\kappa=4$, the channel can be obtained from \eqref{eq10}. We consider that there is a target $d_r=30$ m away from the transmit array. The norm of the complex scattering coefficient $|\alpha|$ of the target is set to $|\alpha|\sim \mathrm{Rayleigh}\left(\sigma_{|\alpha|}\right)$ , where $ \sigma_{|\alpha|}$ can be calculated by $\sqrt{\frac{2}{\pi}\frac{\lambda ^2\boldsymbol{\sigma }_r}{\left( 4\pi \right) ^3d_{r}^{4}}}$ \cite{Skolnik1980}, the radar cross-section (RCS) $ \sigma_r$ is set to 2. The $\theta$ of the target is set to $\theta\sim \mathcal N \left(0,10 \right)$. The false alarm rate is set to ${P_f}=10^{-6}$.
\begin{figure}[t]
	\centering
 \includegraphics[width=0.48\textwidth]{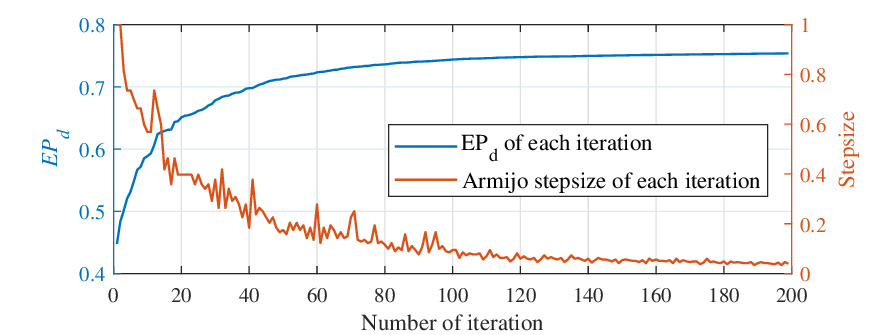}\\	\caption{Expected detection probability and Armijo stepsize versus number of iteration of the proposed Algorithm \ref{Alg1}.}
        \label{Fig2}
\end{figure}
\begin{figure}[t]
\begin{center}
\includegraphics[width=0.49\textwidth]{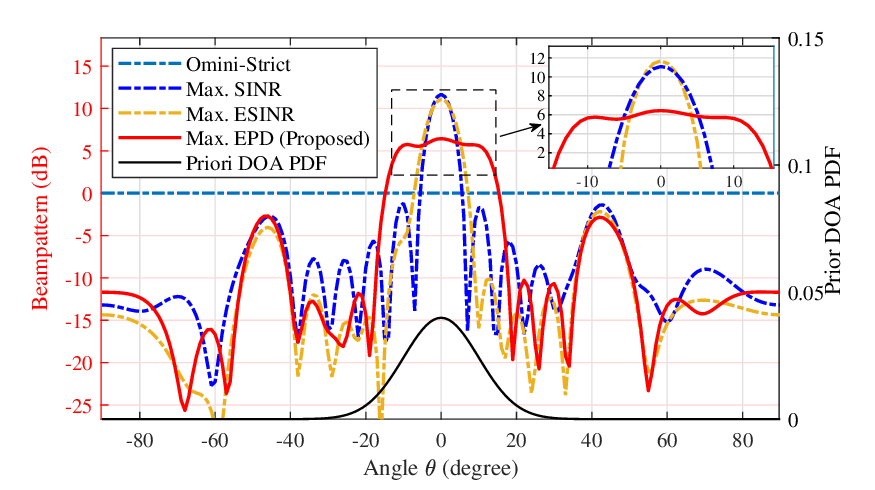}\\	\caption{Beampatterns of various beamforming schemes.}
        \label{Fig3}
\end{center}
\end{figure}

First, we verify the convergence of the proposed Algorithm \ref{Alg1}. The expected detection probability and step size versus the number of iterations are shown in Fig.~\ref{Fig2}. The results in the figure show that the proposed SCA-SDR algorithm can effectively improve the expected detection probability of the target through iteration and has good convergence performance.
\begin{figure}[t]
	\centering
\includegraphics[width=0.49\textwidth]{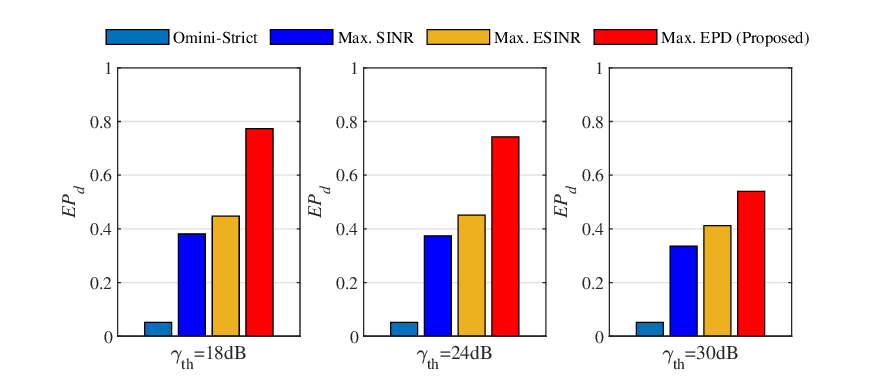}\\	\caption{${EP_d}$ \normalsize of various schemes under different communication threshold.}
        \label{Fig4}
\end{figure}
\begin{figure}[t]
	\centering
\includegraphics[width=0.49\textwidth]{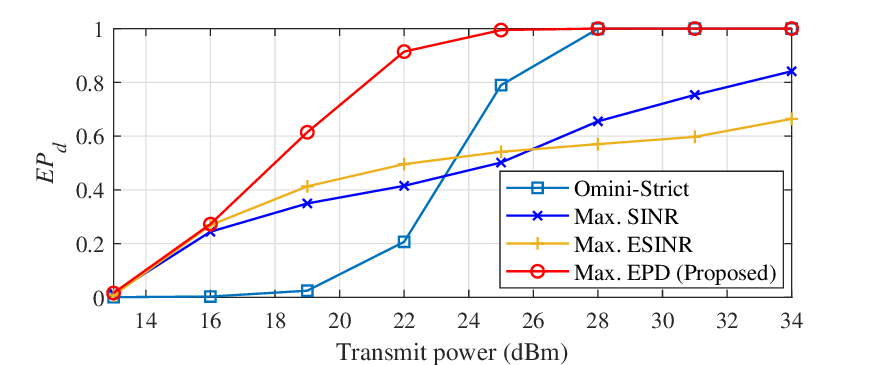}\\	\caption{${EP_d}$ \normalsize of various schemes versus the transmit power}
        \label{Fig5}      
\end{figure}
Moreover, we use several benchmark methods to compare with the proposed scheme (Max. EPD). The Omini Strict and Max. ESINR are schemes proposed in \cite{AttiahX2023}, which are to maximize omnidirectional coverage and maximize expected SINR under energy constraints and communication performance constraints, respectively. The Max. SINR does not consider the uncertainty of the target and maximizes the SINR at $0^{\mathrm{o}}$ under energy constraints and communication performance constraints. Fig.~\ref{Fig3} shows the beampatterns of the proposed scheme and baseline schemes with $\gamma_{\mathrm {th}}=24$ dB. As can be seen from the figure, the proposed scheme can adaptively emit a wider beam to improve the ${EP_d}$ of the target. Fig.~\ref{Fig4} shows the  ${EP_d}$ that can be achieved by different schemes under different communication constraints. It can be clearly seen from the figure that compared with other baseline schemes, our proposed scheme can always achieve the best ${EP_d}$. Fig.~\ref{Fig5} shows the ${EP_d}$ of various schemes under different transmit power. To achieve the same level of ${EP_d}$, the proposed scheme requires significantly smaller transmit power compared to other methods. In addition, when the transmit power is relatively low, the performance of Max. ESINR, Max. SINR, and the proposed scheme are similar. However, as the transmit power increases, the proposed scheme is significantly better than other schemes.

\begin{figure}[ht]
\includegraphics[width=0.49\textwidth]{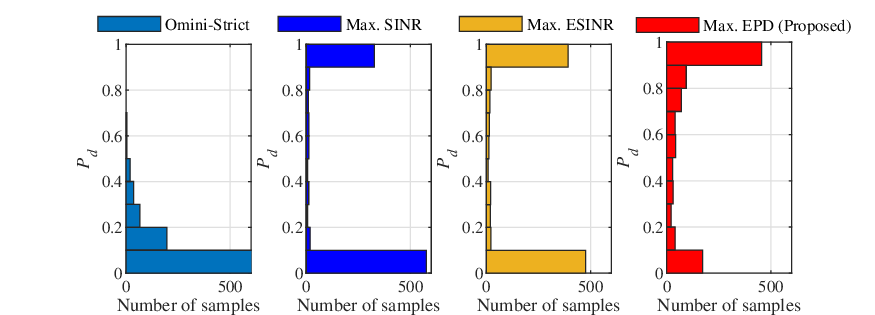}  
 \caption{Detection probability histogram of Monte Carlo samples }
\label{Fig6}      
\end{figure}

Finally, we conduct a Monte Carlo experiment to randomly generate targets according to the distribution of $\alpha$  and $\theta$ and evaluate the detection probability of the targets. The histogram is shown in Fig.~\ref{Fig6}. Fig. 6 shows the histogram of the detection probability for 1000 random samples generated according to the target distribution under different schemes. It can be seen that the detection performance of the proposed scheme is better than other schemes and has strong detection robustness. 

\section{Conclusion}\label{sec5}
To address the challenges posed by target uncertainty, we proposed a Bayesian beamforming scheme for ISAC systems in this paper. We modeled the scattering coefficient and azimuth angle of the target as random variables and defined a new sensing metric of ${EP_d}$. The Bayesian beamforming design problem is formulated as a non-convex optimization problem of maximizing the ${EP_d}$  under the constraints of energy and communication performance. An SCA-SDR algorithm is developed to solve this optimization problem. Simulation results show that the proposed scheme can improve the robust target detection performance and achieves higher detection probability than state-of-the-art methods.

\section*{Acknowledgment}
The work was supported in part by the National Natural Science Foundation of China under Grant 62388102, and the GuangDong Basic and Applied Basic Research Foundation under Grant 2022A1515010209. The corresponding author is Dr. Yuhan Dong.

\appendices
\section{Proof of Proposition 1}
\label{AP1}
We observe the signal $\bm{y}_s$, which under \( H_1 \) follows the model:
\begin{align}
\bm{y}_s=\alpha \bm{V}\left( \theta\right)\bm{x}+\bm{z}_s,
\end{align}
Under \( H_0 \), there is no target, so
\begin{align}
\bm{y}_s=\bm{z}_s,
\end{align}
The GLRT statistic is given by
\begin{align}
{\bm{\Lambda}(\bm{y}_s)} = \max_{\alpha, \theta} \frac{\mathcal{L}_1(\bm{y}_s); \alpha, \theta)}{\mathcal{L}_0(\bm{y}_s)},
\end{align}
where \(\mathcal{L}_1(\mathbf{y}; \alpha, \theta)\) is the likelihood under \( H_1 \),
  \(\mathcal{L}_0(\mathbf{y})\) is the likelihood under \( H_0 \).
\begin{align}
\mathcal{L}_0(\bm{y}_s) = \frac{1}{(\pi \sigma_s^{2})^{(N_rL)}} \exp\left(-\frac{\|\bm{y}_s\|^2}{\sigma_s^2}\right),
\end{align}
Under \( H_1 \), the likelihood is:
\begin{align}
\mathcal{L}_1(\bm{y}_s; \alpha, \theta) = \frac{1}{(\pi \sigma_s^{2})^{(N_rL)}} \exp\left(-\frac{\|\bm{y}_s - \alpha \bm{V}\left( \theta\right)\bm{x}\|^2}{\sigma_s^2}\right).
\end{align}
Substituting the likelihoods into the GLRT definition:
\begin{align}
{\bm{\Lambda}(\bm{y}_s)} = \max_{\alpha, \theta} \frac{\exp\left(-\frac{\|\bm{y}_s - \alpha \bm{V}\left( \theta\right)\bm{x}\|^2}{\sigma_s^2}\right)}{\exp\left(-\frac{\|\bm{y}_s\|^2}{\sigma_s^2}\right)}.
\end{align}
Simplify the exponential terms:
\begin{align}
{\bm{\Lambda}(\bm{y}_s)} = \max_{\alpha, \theta} \exp\left(\frac{-\|\bm{y}_s - \alpha \bm{V}\left( \theta\right)\bm{x}\|^2 + \|\bm{y}_s\|^2}{\sigma_s^2}\right).
\end{align}
Expanding \(\|\bm{y}_s - \alpha \bm{V}\left( \theta\right)\bm{x}\|^2\):
\begin{align}
&\|\bm{y}_s - \alpha \bm{V}\left( \theta\right)\bm{x}\|^2 \nonumber \\
&= \|\bm{y}_s\|^2 - 2 \text{Re}\{\alpha^* \bm{y}_s^H \bm{V}\left( \theta\right)\bm{x}\} + |\alpha|^2 \|\bm{V}\left( \theta\right)\bm{x}\|^2.
\end{align}
Substitute back into ${\bm{\Lambda}(\bm{y}_s)}$:
\begin{align}
{\bm{\Lambda}(\bm{y}_s)} = \max_{\alpha, \theta} \exp\left(\frac{2 \text{Re}\{\alpha^* \bm{y}_s^H \bm{V}\left( \theta\right)\bm{x}\}\} - |\alpha|^2 \|\bm{V}\left( \theta\right)\bm{x}\|^2}{\sigma_s^2}\right).
\end{align}
To find the optimal $\hat{\alpha}$, maximize the argument of the exponential with respect to \(\alpha\),
\begin{align}
\hat{\alpha} = \frac{\bm{y}_s^H \bm{V}\left( \theta\right)\bm{x}}{\|\bm{V}\left( \theta\right)\bm{x}\|^2}.
\end{align}
Substitute $\alpha$ back into $\bm\Lambda(\bm{y}_s)$,
\begin{align}
\bm\Lambda(\bm{y}_s) = \max_{\theta} \exp\left(\frac{\left|\bm{y}_s^H \bm{V}\left( \theta\right)\bm{x}\right|^2}{\sigma_s^2 \|\bm{V}\left( \theta\right)\bm{x}\|^2}\right).
\end{align}
Taking the logarithm for simplicity:
\begin{align}
\ln \bm\Lambda(\bm{y}_s) = \max_{\theta} \frac{\left|\bm{y}_s^H \bm{V}\left( \theta\right)\bm{x}\right|^2}{\sigma_s^2 \| \bm{V}\left( \theta\right)\bm{x}\|^2}.
\end{align}
The GLRT detector can be formulated as
\begin{align}
\max_{\theta} \frac{\left|\bm{y}_s^H \bm{V}\left( \theta\right)\bm{x}\right|^2}{\|\bm{V}\left( \theta\right)\bm{x}\|^2} \begin{array}{c}
	\overset{\mathcal{H}_1}{\geqslant}\\
	\underset{\mathcal{H}_0}{<}\\
\end{array}\,\, \varGamma,
\end{align}
where $\varGamma$ is a threshold determined by the desired false alarm probability.

\balance 


\begin{thebibliography}{00}
\bibitem{Hassan2016} A. Hassanien, M. G. Amin, Y. D. Zhang, and F. Ahmad, ``Signaling strategies for dual-function radar communications: An overview,'' {\em IEEE Aerosp. Electron. Syst. Mag.}, vol. 31, no. 10, pp. 36--45, Oct. 2016.

\bibitem{Liu2022} F. Liu, Y. Cui, C. Masouros, J. Xu, T. X. Han, Y. C. Eldar, and S. Buzzi, ``Integrated sensing and communications: Towards dual-functional wireless networks for 6G and beyond,'' {\em IEEE J. Sel. Areas Commun.}, vol. 40, no. 6, pp. 1728--1767, Jun. 2022.

\bibitem{Cui2021} Y. Cui, F. Liu, X. Jing and J. Mu, ``Integrating sensing and communications for ubiquitous IoT: Applications, trends, and challenges,'' {\em IEEE Network}, vol. 35, no. 5, pp. 158-167, Nov. 2021.

\bibitem{Zhao2022} Z. Zhao, X. Tang, and Y. Dong, ``Cognitive waveform design for dual-functional MIMO radar-communication systems,'' in {\em Proc. IEEE Global Commun. Conf. (GLOBECOM)}, Dec. 2022, pp. 5607--5612.

\bibitem{LiuF2024} F. Liu, T. Zhang, Z. Zhang, Y. Shen and Q. Zhang, ``ISAC with UWB: Reliable decoupling and target sensing,'' {\em IEEE Trans. Wireless Commun.}, 2024. (early access) 

\bibitem{Kwon2023} G. Kwon, Z. Liu, A. Conti, H. Park, and M. Z. Win, ``Integrated localization and communication for efficient millimeter wave networks,'' {\em IEEE J. Sel. Areas Commun.}, vol. 41, no. 12, pp. 3925--3941, Dec. 2023.

\bibitem{ZhaoZ2024} Z. Zhao, T. Wei, Z. Liu, X. Tang, X-P Zhang, and Y. Dong, ``Joint beamforming for backscatter integrated sensing and communication,'' arXiv:2409.02797.

\bibitem{LiuX2020} X. Liu, T. Huang, N. Shlezinger, Y. Liu, J. Zhou, and Y. C. Eldar, ``Joint transmit beamforming for multiuser MIMO communications and MIMO radar,'' {\em IEEE Trans. Signal Process.}, vol. 68, pp. 3929--3944, Jun. 2020.

\bibitem{LiuF2022} F. Liu, Y. -F. Liu, A. Li, C. Masouros and Y. C. Eldar, ``Cramér-Rao bound optimization for joint radar-communication beamforming,'' {\em IEEE Trans. Signal Process.}, vol. 70, pp. 240-253, Dec. 2022.

\bibitem{Zhao2024} Z. Zhao, {\em et al.}, ``Joint beamforming scheme for ISAC systems via robust Cramér–Rao bound optimization,'' {\em IEEE Wireless Commun. Lett.}, vol. 13, no. 3,  pp. 889--893, Jan. 2024.

\bibitem{Zhao2025} Z. Zhao, Z. Liu, R. Jiang, Z. Li, X.-P. Zhang, X. Tang, and Y. Dong ``Joint beamforming for multi-target detection and multi-user communication in ISAC systems,'' arXiv:2411.00433.

\bibitem{AttiahX2023} K. M. Attiah and W. Yu,, ``Active beamforming for integrated sensing and communication,'' {\em 2023 IEEE International Conference on Communications Workshops (ICC Workshops)}, 2023, pp. 1469-1474.

\bibitem{Xu2024} C. Xu and S. Zhang, ``MIMO integrated sensing and communication exploiting prior information,'' {\em IEEE J. Sel. Areas Commun.}, vol. 42, no. 9, pp. 2306-2321, Sept. 2024.

\bibitem{Hou2024}K. Hou and S. Zhang, ``Optimal beamforming for secure integrated sensing and communication exploiting target location distribution,'' {\em IEEE J. Sel. Areas Commun.}, vol. 42, no. 11, pp. 3125-3139, Nov. 2024.

\bibitem{Tse2005} D. Tse and P. Viswanath, {\em Fundamentals of Wireless Communication.}, Cambridge, U.K.: Cambridge Univ. Press, 2005.

\bibitem{Fortu2020} S. Fortunati, L. Sanguinetti, F. Gini, M. S. Greco, and B. Himed, ``Massive MIMO radar for target detection,'' {\em IEEE Trans. Signal Process.}, vol. 68, pp. 859-871, Jan. 2020.

\bibitem{Neyman1992} J. Neyman and E. S. Pearson, ``On the problem of the most efficient tests of statistical hypotheses,'' {\em Philosophical Trans. the Royal Society of London}, vol. 231, no. 694-706, pp. 289--337, 1992.

\bibitem{Kay1998}S. M. Kay, {\em Fundamentals of Statistical Signal Processing-Volume II: Detection Theory}, Englewood Cliffs, NJ, USA: Prentice Hall, 1998.

\bibitem{Tang2022} B. Tang and P. Stoica, ``MIMO multifunction RF systems: Detection performance and waveform design,'' {\em IEEE Trans. Signal Process.}, vol. 70, pp. 4381-4394, Aug. 2022.

\bibitem{boyed2004}S.Boyd and L. Vandenberghe, {\em Convex Optimization.}, Cambridge University Press, 2004.

\bibitem{Luo2010} Z.-Q. Luo, W.-K. Ma, A. M.-C. So, Y. Ye, and S. Zhang, ``Semidefinite relaxation of quadratic optimization problems,'' {\em IEEE Signal Process. Mag.}, vol. 27, no. 3, pp. 20--34, May 2010.

\bibitem{Scutari2014} G. Scutari, F. Facchinei, P. Song, D. P. Palomar and J. S. Pang, ``Decomposition by partial linearization: Parallel optimization of multi-agent systems,'' {\em IEEE Trans. Signal Process.}, vol. 62, no. 3, pp. 641--656, Feb. 2014.

\bibitem{Skolnik1980} M. I. Skolnik, {\em Introduction to Radar Systems.}, vol. 3. New York, NY, USA: McGraw-Hill, 1980.

\bibliographystyle{IEEEtran}
\end{thebibliography}
\end{document}